\pdfoutput=1
\documentclass[aps,prb,showpacs,preprint]{revtex4}

\usepackage{amssymb}
\usepackage{amsmath}
\usepackage{graphicx}

\begin{document}

\title{Toward equilibrium ground state of charge density waves in rare-earth tritellurides }
\date{\today}

\author{A.V.~Frolov$^1$, A.P.~Orlov$^{1,2}$, A.~Hadj-Azzem$^{3}$, P.~Lejay$^{3}$, A.A.~Sinchenko$^{1,4}$, and P.~Monceau$^{3}$}

\address{$^{1}$Kotelnikov Institute of Radioengineering and Electronics of RAS, 125009 Moscow, Russia}

\address{$^{2}$Institute of Nanotechnologies of Microelectronics of RAS, 115487, Moscow, Russia}

\address{$^{3}$Univ. Grenoble Alpes, Inst. Neel, F-38042 Grenoble, France, CNRS, Inst. Neel, F-38042 Grenoble, France}

\address{$^{4}$M.V. Lomonosov Moscow State University, 119991, Moscow, Russia}

\begin{abstract}

We show that the charge density wave (CDW) ground state below the Peierls transition temperature, $T_{CDW}$, of rare-earth tritellurides is not at its equilibrium value, but depends on the time where the system was kept at a fixed temperature below $T_{CDW}$. This ergodicity breaking is revealed by the increase of the threshold electric field for CDW sliding which depends exponentially on time. We tentatively explain this behavior by the reorganization of the oligomeric (Te$_x$)$^{2-}$ sequence forming the CDW modulation.

\end{abstract}


\maketitle

\section{Introduction}

The collective motion (sliding) of charge density waves (CDW) is one of the most fascinating property of low-dimensional
compounds exhibiting this type of electronic ordering. First predicted by Fr\"{o}hlich \cite{Frohlich54} as dissipationless
charge transfer, this type of electron transport is possible only when the electric field reaches a threshold electric field, $E_t$. Impurities, defects, interchain interaction or commensurability with the main lattice pin the phase of the CDW, and the conductivity in low electric fields, $E<E_t$, demonstrates the conventional ohmic behavior because of quasiparticle excitations. At $E>E_t$ the pinning is overcome and the CDW starts to slide that is manifested as a sharp increase of conductivity. Such type of transport was previously observed and well studied in many inorganic and organic quasi-one-dimensional compounds \cite{Gruner,Monceau12}.

Fr\"{o}hlich considered the CDW as a macroscopic quantum state and suggested that there can be states with current flow if the CDW energy gap is displaced with the electrons and remains attached to the Fermi surface. This model was revised in Refs. \onlinecite{Bardeen74,Bardeen85,Bardeen89}. If the electron dispersion is displaced by $q$, that leads a Fr\"{o}hlich collective current $j_{CDW}=nev$ with $\hbar q=m^*v$ where $m^*$ is the Fr\"{o}hlich CDW mass. The Fr\"{o}hlich model captures very well the CDW non-linear transport properties \cite{Monceau12}. In another electronic crystal, namely La$_{1.875}$Ba$_{0.125}$CuO$_4$ with charge ordering (CO), a shift $q$ in momentum of the CO peak was measured from femtosecond resonant X-ray scattering which can be interpreted as the CO condensate moving with momentum $q$ following the Fr\"{o}hlich model \cite{Mitrano19}.

Recently, CDW sliding was also discovered in quasi two-dimensional (2D) rare-earth tritellurides RTe$_3$ (R: La, Ce, Pr, Nd, Gd, Tb, Dy, Er, Tm) compounds \cite{SinchPRB12,SSC14}. RTe$_3$ family have an orthorhombic crystal structure (space group $C_{mcm}$) and consist of double Te planes separated by corrugated RTe planes. In this space group, the $b$-axis is perpendicular to Te planes. All these compounds exhibit a transition to a state with an incommensurate unidirectional charge density wave with a wave vector ${\bf Q}_{CDW1}=(0,0;\sim\dfrac{2}{7}c^*)$ at the Peierls temperature higher than 250 K. For heavy R elements (Dy, Ho, Er, Tm), a second transition to a state with an independent charge density wave with the wave vector ${\bf Q}_{CDW2}=(\sim\dfrac{2}{7}a^*,0,0)$ perpendicular to ${\bf Q}_{CDW1}$ occurs at low temperatures. The sharpness of superstructure maxima in X-ray diffraction indicates the existence of a three-dimensional long-range order \cite{Ru08}.

In this paper, we report a new effect in the non-linear CDW dynamic observed in TbTe$_3$. The threshold electric field increases dramatically under isothermal exposition of the sample at a fixed $T_{exp}$ below $T_{CDW}$ during several hours. The temperature dependence of $E_t$ after such procedure demonstrates a strong peak effect with a maximum at $T_{exp}$. Some similarity can be found with the peak effect in superconductors \cite{Larkin79}. These effects have never been observed in quasi one-dimensional (1D) CDW systems.

\section{Experimental}

\begin{figure}[tbh]
	\includegraphics[width=8.5cm]{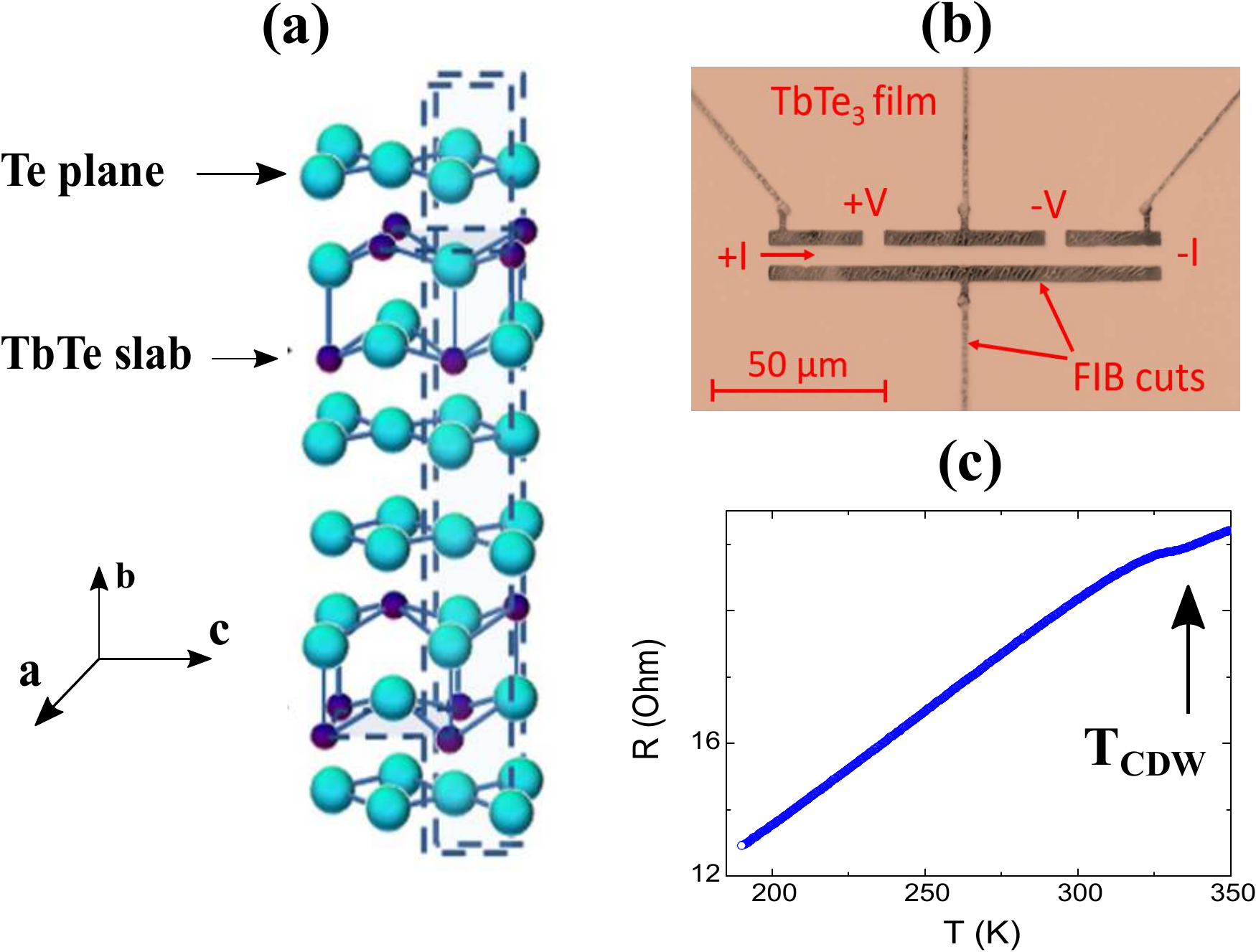}
	\caption{(a) Sketch of the TbTe$_3$ structure. (b) Optical microscope image of the typical
		bar structure based on a TbTe$_3$ single crystal. (c) Typical temperature dependence of resistance, $R(T)$, for one of the structure.}
	\label{F0}
\end{figure}

Fig. \ref{F0} (a) shows a sketch of the crystal structure of TbTe$_3$. The lattice parameters of TbTe$_3$ are $a=4.29$ \AA, $b=25.33$ \AA  and $c=4.303$ \AA. A CDW with a Peierls transition at $T_{CDW}=336$ K develops along the $c$-axis with the wave vector ${\bf Q}_{CDW1}=(0,0,0.296)$. A CDW has also been reported \cite{BB} below $T\sim 41$ K along the $a$-axis with a wave vector ${\bf Q}_{CDW2}=(0.32,0,0)$. Although the in-plane lattice parameters are almost identical, the Te planes are essentially square with the C$_4$ symmetry. As shown by Yao et al. in Ref. \onlinecite{Yao06}, a CDW in a layered quasi 2D system with a tetragonal symmetry can either bidirectional (checkboard) or unidirectional. They derived a phase diagram as a function of the electron-phonon coupling. With a strong electron coupling the CDW is unidirectional as it is the case in RTe$_3$ compounds for which the structure is fundamentally orthorhombic because the glide plane between the adjacent Te layers \cite{Eiter13}. Single crystals of this compound were grown in a pure argon atmosphere by the technique described in our previous work \cite{SinchPRB12}. Thin rectangular single-crystal samples thinner than 1 $\mu$m were prepared by the micromechanical exfoliation of relatively thick crystals preliminarily glued to a sapphire substrate. The quality of crystals and the spatial position of crystallographic axes were monitored by X-ray diffraction. Bar structures oriented along $c$-axis with a length of $70-150$ $\mu$m and a width of $5-10$ $\mu$m were prepared by etching with the use of focused ion beam (Fig.\ref{F0} (b)). One of such a structure is shown in  Fig.\ref{F0} (b). Resistances, current-voltage characteristics (IVc) and their derivatives, $dV/dI$, of these structures were measured by the four-probe method. Typical temperature dependence of resistance for one of a structure is shown in  Fig.\ref{F0} (c). It can be seen that the resistivity peculiarity corresponding to the CDW transition at $T=336$ K is very weak that is typical for resistivity along c-axis direction that is in accordance of results of Ref. \onlinecite{anis}. Time evolution of IVc at stabilized temperature was measured automatically with a fixed time interval, typically 30-60 minutes, during several tens of hours.

\section{Experimental results}

\begin{figure}[tbh]
	\includegraphics[width=8.5cm]{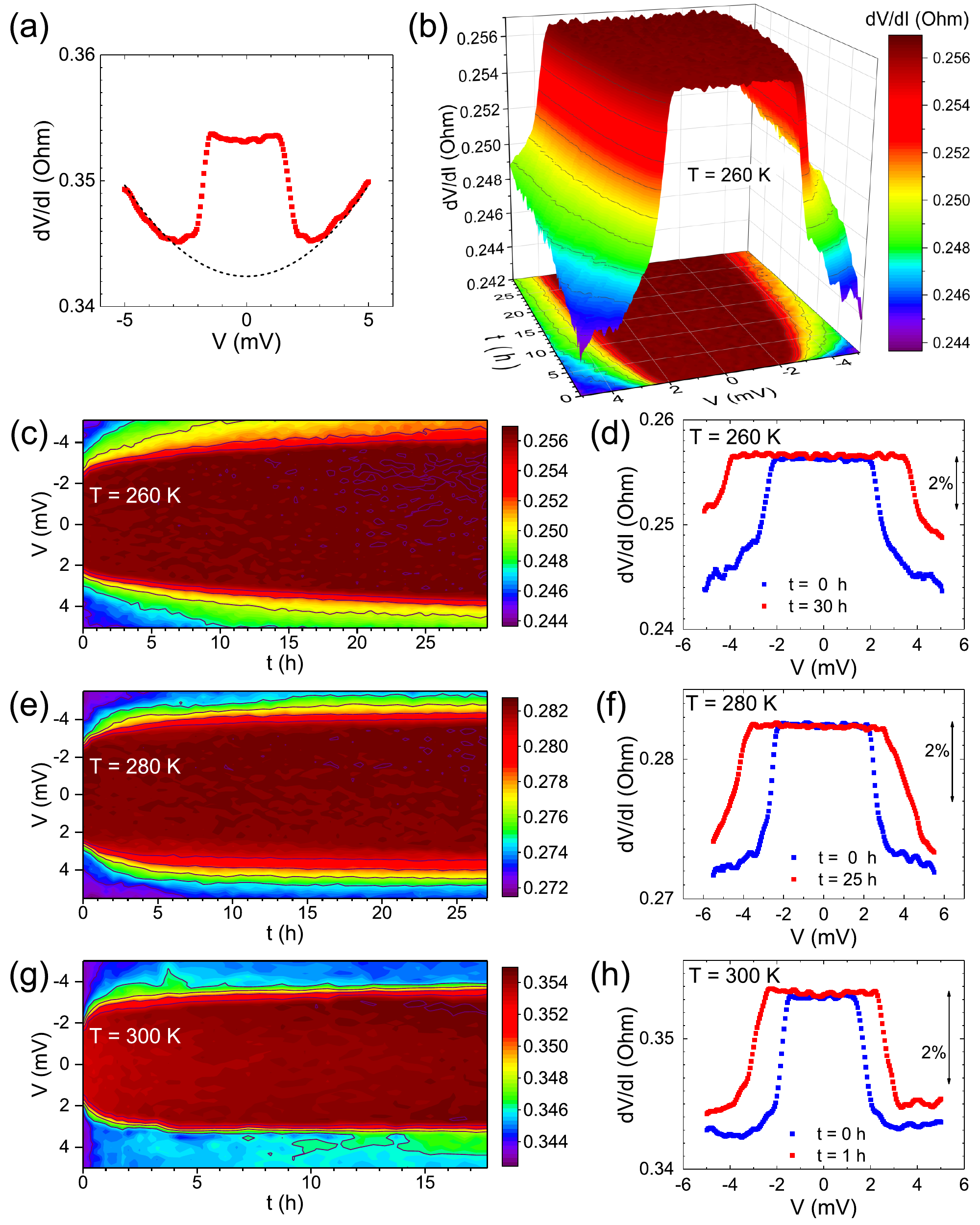}
	\caption{(a) Original differential current–-voltage characteristic, $dV/dI(V)$, at $T=260$ K. Dashed line shows the parabolic background corresponding to Joule heating; (b) Three-dimensional plot of time evolution of differential IVs after subtraction of Joule heating; (c), (e), (g) Evolution of differential IVc with time for $T_{exp}=$260; 280 and 300 K; (d), (f), (h) initial (blue) and final (red) differential IVs with substraction of Joule heating for the same temperatures.}
	\label{F1}
\end{figure}

In usual experimental condition when the TbTe$_3$ sample is cooled from $T>T_{CDW}$ at the rate 3-5 K/min the temperature dependence of the threshold electric field, $E_t(T)$, of the order of $10^2$ mV/cm, reproduces the behavior previously reported in Ref. \onlinecite{SSC14}, that is to say a nearly linear increase when $T$ is decreased. However, we noticed that $E_t$ increases significantly if the sample is kept at a fixed $T_{exp}$ below $T_{CDW1}$ a sufficiently long time \cite{PE19}. To study this effect the experimental procedure was the following: we cooled the sample from $T>T_{CDW1}$ down to a given temperature $T_{exp}$ and measured IVc with a time interval of 30 minutes during several tens of hours and that at different $T_{exp}$ in the range 220-330 K. Before each exposition the sample was warmed above $T_{CDW1}$. For more clarity we extracted the parabolic contribution of Joule heating (dashed line in Fig. \ref{F1} (a)) from original IVs. A three dimensional plot illustrating  measurements is shown in Figs. \ref{F1} (b) for $T_{exp}=260$ K. In Figures (c), (e) and (g) we show the evolution of differential IVc with time for one of the samples for $T_{exp}=$ 260, 280 and 300 K correspondingly. Figs. \ref{F1} (d), (f) and (h) represent initial and final $dV/dI(V)$ for corresponding $T_{exp}$. It can be seen that the threshold electric field really increases with time and saturates at a certain value of $t$. This characteristic time depends on the chosen $T_{exp}$. This time evolution of the threshold is more pronounced for higher values of $T_{exp}$ for which $E_t$ increases more than two times.

It is seen in Figs. \ref{F1} (d, f, h) that, in addition in the change value of $E_t$, the drop of resistance at the threshold which determines the contribution of the sliding of the CDW to electron transport, is significantly less at low $T_{exp}$. The contribution of the collective motion of the CDW to the total transport current is determined by the number of carriers condensed in the CDW state and the velocity of the CDW. In our case, the number of carriers condensed below the CDW gap should be the same and not be depending on the thermal treatment at $T_{exp}$. Consequently, the sliding velocity of the CDW determined by friction effects for exposed samples is much lower. Note, that the resistance at $V=0$ remains the same \cite{PE19} independently of the exposition time at all $T_{exp}$ as it seen in Figs. \ref{F1} (d, f, h). The resistance in the CDW static state is determined by carrier scattering on the Fermi surface reduced by the opening of the CDW energy gap which does not change with time evolution at any $T_{exp}$. Thus, this result is the direct indication that all the changes in the sample during the exposition take place in the CDW subsystem.

\begin{figure*}[tbh]
	\includegraphics[width=0.3\textwidth]{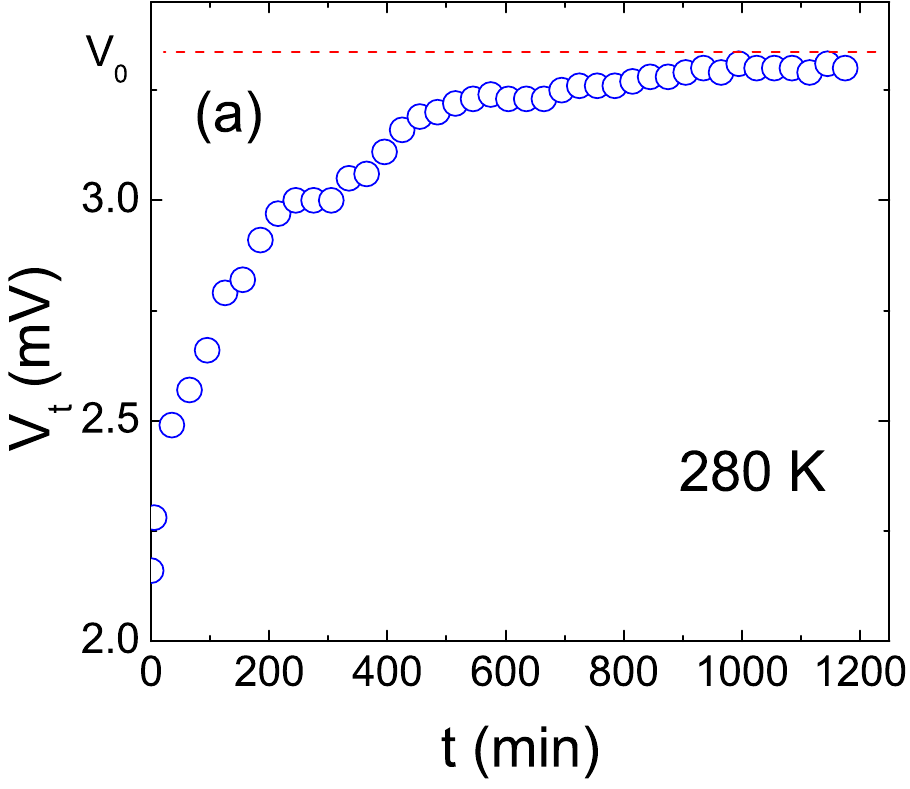} \ \
	\includegraphics[width=0.3\textwidth]{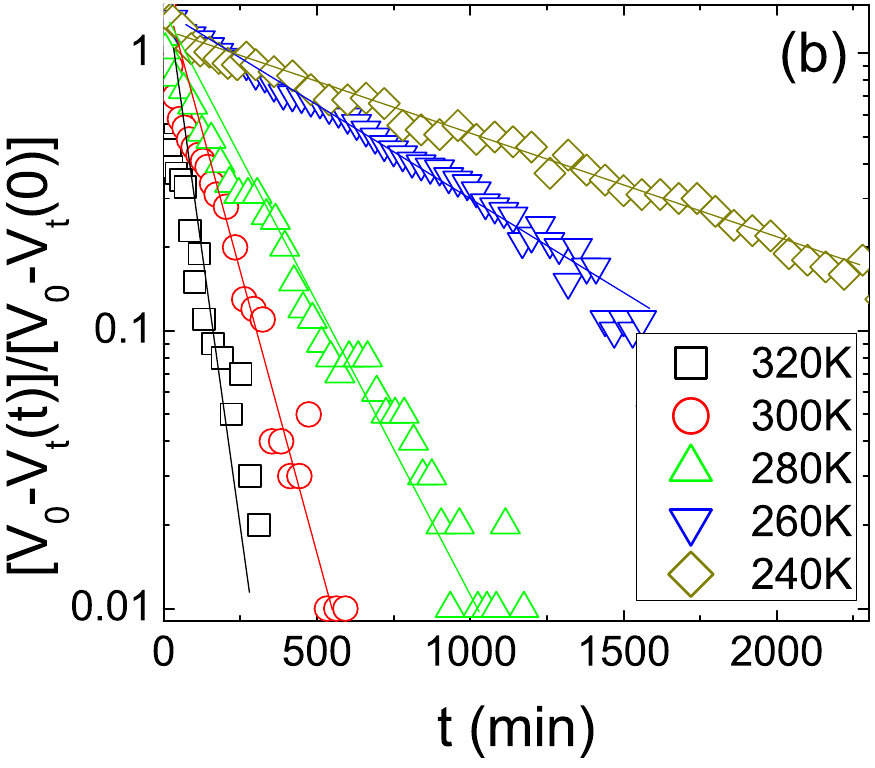} \ \
	\includegraphics[width=0.3\textwidth]{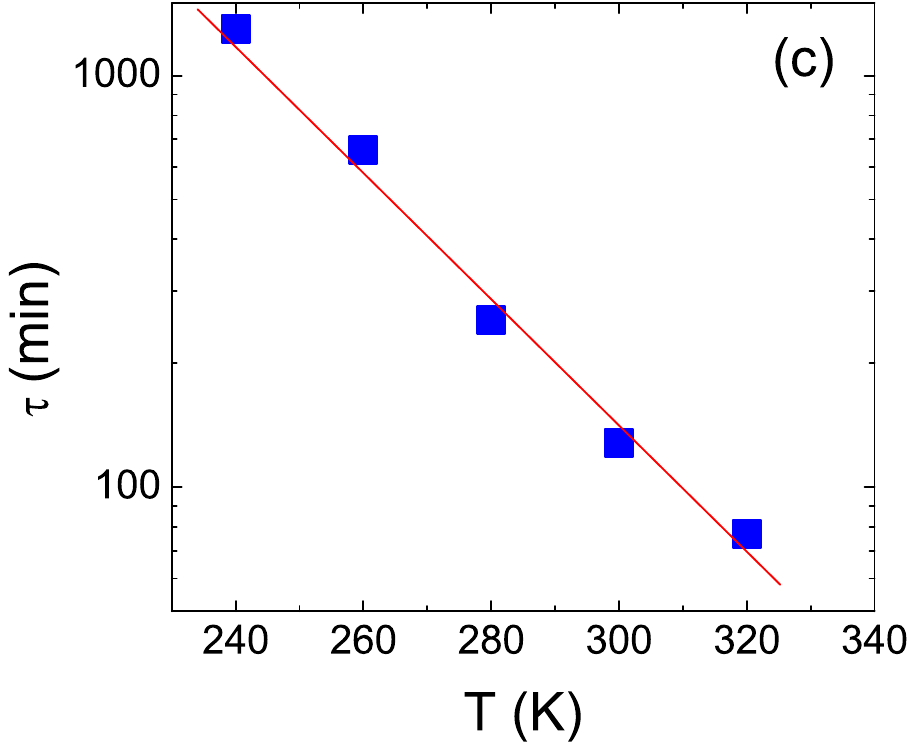}
	\caption{a) Dependence $V_t(t)$, where $V_t$ is the threshold voltage, at $T_{exp}=280$ K; (b) semi-logarithmic plot of $[V_0-V_t(t)]/[V_0-V_t(t=0)]$ for $T_{exp}=$ 320 K (black squares); 300 K (red circles); 280 K (green triangles); 260 K (blue inverted triangles) and 240 K (gray rhombuses); (c) Temperature dependence of the relaxation time, $\tau$, deduced from (b) as a function of temperature in a semi-logarithmic plot.} 
	\label{F2}
\end{figure*}

Fig. \ref{F2} (a) shows the dependence of $V_t(t)$, where $V_t$ is the threshold voltage, at $T_{exp}=280$ K. As can be seen, the threshold voltage, $V_t(t)$ rapidly increases at short time and saturates at a value $V_0$ at larger time. To determine the characteristic time of the $V_t(t)$ evolution, we have plotted in Fig. \ref{F2} (b) the dependencies of $[V_0-V_t(t)]/[V_0-V_t(t=0)]$ for different $T_{exp}$ in logarithmic scale. At all temperatures $(V_0-V_t(t))\sim(1-exp(t/\tau))$ with the characteristic time $\tau$ strongly temperature dependent. As can be seen from Fig. \ref{F2} (c) $\tau$ depends exponentially on temperature as $\tau\sim exp(-\frac{T}{T_0})$, with $T_0\approx30$ K.

\begin{figure}[tbh]
	\includegraphics[width=8.5cm]{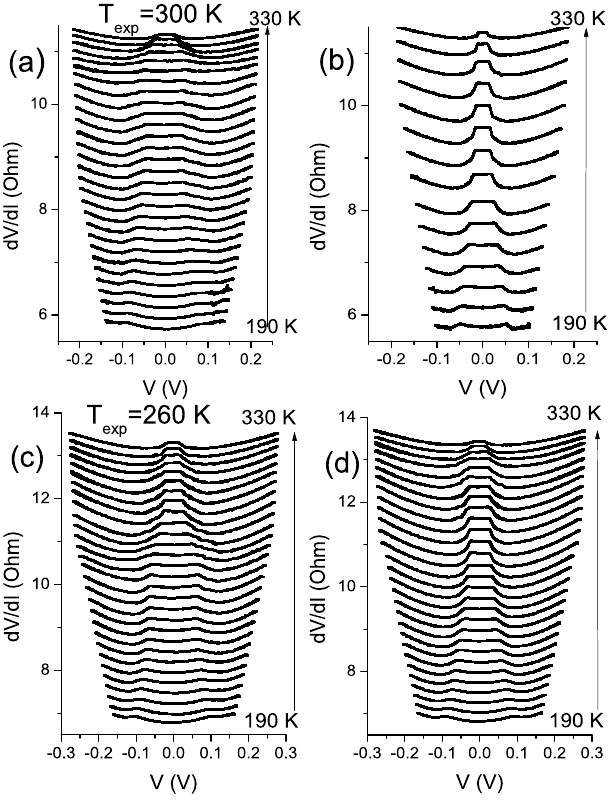}
	\caption{(a) Differential IVs measured under continuous warming in a temperature range 190-330 K with a step $\Delta T=5$ K of TbTe$_3$ exposed at $T_{exp}=300$ K during 20 hours. (b) Differential current–-voltage characteristics measured in the same temperature range with a step $\Delta T=10$ K for the same sample during continuous cooling from $T=350$ K. (c) The same as in a) but for $T_{exp}=260$ K. (d) Differential IVc during continuous cooling from $T=330$ K after warming in (c).}
	\label{F3}
\end{figure}

\begin{figure}[tbh]
	\includegraphics[width=8.5cm]{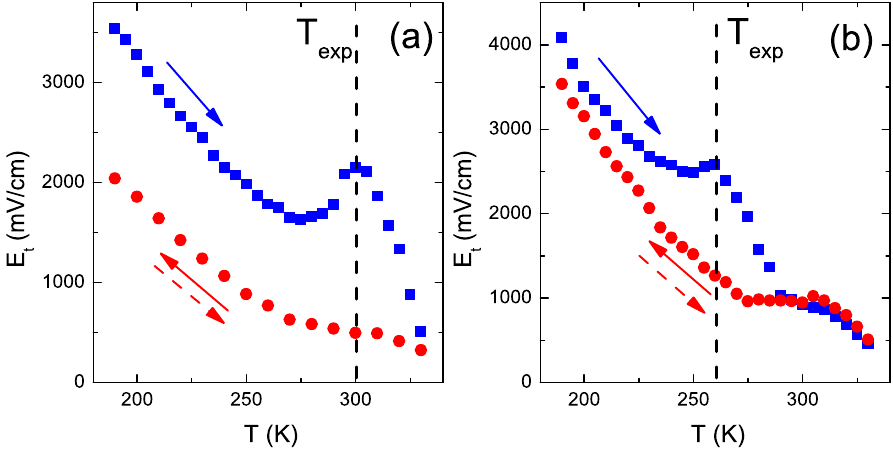}
	\caption{Temperature dependence of the threshold electric field, $E_t$, for the CDW sliding for exposed samples (blue squares) and not-exposed (red circles) for $T_{exp}=300$ K (a) and 260 K (b). Red dots correspond to $E_t$ when the sample is continuously cooled at a rate of a few degrees per minute from $T>T_{CDW1}$ down to 190 K. Blue squares are values of $E_t$ measured in warming when the sample was exposed at $T_{exp}=300$ K (a) and 260 K (b) for a long time/ then cooled to 190 K. The maximum of $E_t$ occurs at the temperature of exposition, $T_{exp}$, as shown by dashed lines in (a) and (b).}
	\label{F4}
\end{figure}

Let us consider the temperature dependence of the threshold electric field. As it was reported in Ref. \onlinecite{SSC14}, when the temperature is continuously decreased from $T$ well above $T_{CDW1}=336$ K the temperature of the threshold electric field demonstrates a nearly linear increase. For samples not exposed at $T_{exp}$ for a given time, the dependence $E_t(T)$ is completely reversible as indicated in fig. \ref{F4} by arrows with both directions. Another picture is observed for samples which were exposed a long time at a fixed $T_{exp}$. These samples were cooled from $T>T_{CDW1}$ down to $T_{exp}$, kept at this temperature for a given time and then cooled down to $T=190$ K with a rate 5 K/min, then IV curves were measured under warming with a fixed temperature step $\Delta T$. For each step the temperature was regulated with an accuracy of 0.1 K. Fig. \ref{F3} (a) shows a set of IVc in the temperature range 190-330 K for a sample which was exposed during 20 hours at $T=300$ K. Fig. \ref{F3} (b) shows IV curves for the same sample but without exposition, the measurements have been performed similarly as in Ref. \onlinecite{SSC14}. It is seen that the characteristics in Fig. \ref{F3} (b) are in agreement with known results\cite{SSC14}: the threshold field increases monotonically when $T$ is decreased, demonstrating a nearly linear dependence. The picture changes qualitatively when the sample is exposed at 300 K. The threshold field increases significantly in the entire temperature interval, excluding a narrow range near $T_{CDW}$. The dependence $E_t(T)$ becomes nonmonotonic with a maximum near $T=T_{exp}$ (Fig. \ref{F3} (a)).

Qualitatively the same picture is observed for different $T_{exp}$. As an example, Figs. \ref{F3} (c) and (d) demonstrate corresponding temperature evolutions of differential IVs: under warming from 190 K up to 330 K for a sample which was exposed during 20 hours at $T_{exp}=260$ K  (c) and under continuous cooling the same sample from 330 K up to 190 K (d). The curves in Fig. \ref{F3} (d) show a nearly linear increase of $E_t$ when $T$ is decreased while curves in Fig. \ref{F3} (c) show a strong maximum near $T=T_{exp}=260$ K.

One question that has to be considered: is the effect on the value of $E_t$ is depending on the time where the sample is in the non linear state. Let’s recall the experimental procedure: at $T_{exp}$,  IVc are measured every half an hour and the time for recording each IVc approximately 1 minute. Thus, the time at which the voltage is above the threshold is two orders of magnitude with respect to that in the static state. Then we conclude that the time spent at $T_{exp}$ which determines the evolution of $E_t$. Temperature dependencies of $E_t(T)$ for both cases shown in Fig. \ref{F3}, for $T_{exp}=300$ K (a) and 260 K (b), are drawn in Fig. \ref{F4}. Red circles correspond to $E_t$ values measured when $T$ is continuously decreased from $T>T_{CDW1}$  for Fig. \ref{F4} (a) or continuously decreased from $T=330$ K  for Fig. \ref{F4} (b). Blue squares correspond to $E_t$ values when, the temperature being reduced from $T>T_{CDW1}$ and kept fixed at $T=T_{exp}<T_{CDW}$ for enough time to reach saturation as shown in Fig. \ref{F2}, $T$ is continuously increased from $T=190$ K, manifesting sharp maxima at $T\sim T_{exp}$. Note that the $E_t(T)$ dependencies without or with exposition at $T=260$ K become identical 30-40 K above $T=T_{exp}$ as it seen in Fig. \ref{F4} (b).

\section{Discussion}

To our knowledge such time effect on the value of the threshold field $E_t$ has never been reported earlier for the CDW compounds exhibiting sliding. Corresponding as mentioned above that the ohmic resistivity of the samples remains unchanged during exposition (see Fig. \ref{F1} (d, f, h), it means that there are no lattice defect created during the exposition times which may act as new pinning centers. So, the strong increase of $E_t$ with time should result from some modification of pinning conditions: appearance of a stronger pinning when the CDW tends to equilibrium ground state.

Let us remind that two types of pinning are usually distinguished: collective or weak pinning where, from an internal random summation of the pinning forces from over a great number of impurities in a domain, the CDW phase deviation to ideality is in the order of $\pi$, and the local or strong pinning where the CDW phase is located at each impurity and produces local metastable states - plastic deformation or dislocation loops - with finite barriers. Competition between weak and strong pinning was perfectly illustrated in the explanation of the low frequency peak in the real part of the dielectric response \cite{LB95}. With the decrease of temperature because of the reduction of screening and exponentially vanishing of the number of free carriers, the weak pinning is exhausted; the pinning becomes essentially local induced by strong pinning impurities which only are effective for initiating plastic deformations and metastable states in the CDW superstructure, thus dominating the kinetic properties of the system. These low energy excitations (LEE) contribute also at temperatures below 1 K in thermodynamical properties as an additional contribution to the specific heat of the regular phonon term. These LEE were interpreted as metastable states and analyzed as two-level systems resulting from local deformation of the CDW at strong pinning impurity centers \cite{Biljakovic89}. In this low-$T$ range the relaxation of these metastable states has the form of a stretched exponential reflecting the wide time distribution. The breakdown of the collective pinning and the crossover to strong pinning of vortices has also been proposed for the explanation of the "peak effect", namely the maximum of the critical current as a function of magnetic field or temperature observed in low- and high-temperature superconductors \cite{Marley95,Kwok94}.

Collective electron transport in RTe$_3$ compounds is possible only along the direction of the CDW wave vector \cite{SSC14}. Although having all signatures of the sliding of a charge density wave, there are significant differences with the CDW sliding in quasi-1D systems. i) the coherent X-ray diffraction pattern in TbTe$_3$ remains nearly identical to the same distribution of speckles up to the threshold, unlike to the 1D NbSe$_3$ in which the CDW is strongly deformed below $E_t$; ii) at $E_t$ the CDW is suddenly depinned and the $2k_F$ wave vector, $Q_{CDW1}$, slightly rotates with atypical angle of $0.002^\circ$, while no compression or dilatation is observed for the component of $Q_{CDW1}$ along the $c$-axis. There is a CDW shear manifested by small component of $Q_{CDW1}$ along $a$- and $b$-axis transverse to the CDW wave vector \cite{Bolloch16,Bellec}; iii) an anomalously small temperature-independent contribution of the CDW motion to the total electron transport determined by the ratio $\Delta R/R$ ($\Delta R$ is the relative change in the resistance at sliding and $R$ is the total resistance), which indicates a very low velocity of this motion in a given electric field; (iv) a linear dependence of $E_t(T)$ \cite{SSC14}. It is worth to note that the threshold field in quasi-1D compounds increases exponentially below $T_{CDW}$ \cite{Maki86}. These results indicate a fundamental difference in sliding mechanisms and, therefore, in pinning mechanisms in 1D and in 2D compounds.

Tellurium being the least electronegative element among chalcogens can stabilize longer than normal bond distances which associate through Te-Te bonding interaction: from the superspace crystallographic technique, it was shown that the CDW modulation in RTe$_3$ compounds can be defined as a sequence in the Te planes of oligomeric fragments of (Te$_x$)$^{2-}$, typically trimers or tetramers, and even single Te atom. For instance the detailed temperature dependence of the incommensurate CDW modulation of SmTe$_3$ can be described as such of sequence of trimers and tetramers. As the temperature is increased the distribution of Te-Te bonds changes and the connectivity of the oligomers now result in a different sequence of trimers and tetramers \cite{Malliakas06,Kim06}. 

Then one can interpret the time dependence of the threshold field as follows: cooling TbTe$_3$ from $T>T_{CDW1}$ at a uniform rate the sequence of oligomers which defines the CDW modulation is metastable with many defects. Keeping the temperature fixed at $T_{exp}$, the oligomeric sequence reorganizes and reaches after a certain time, longer at low temperature, the equilibrium state. A possible mechanism is hopping of Te between trimer and tetramer fragments or the participation of single Te atom in bonding. At the equilibrium the CDW superstructure is more ordered with domains the walls of which can pin the CDW phase collectively, then increasing the threshold field. We assume that under increase of $T$ above $T_{exp}$ the sequence of oligomers becomes metastable again demonstrating some kind of melting. As a result, the crossover from weak collective pinning to strong individual pinning takes place leading to a strong increase of $E_t$ and the appearance of a sharp maximum in the $E_t(T)$ dependence. In the frame of this scenario the processes of ordering/disordering of the oligomer sequence are operated by the thermal fluctuations.

This effect of reorganization of the CDW with time should be observable in high resolution X-ray diffraction since only small change either in the modulation wave vector and/or of the width of the satellite peak are expected. Our X-ray experiment in Ref. \onlinecite{Bolloch16} was performed at room temperature where the CDW satellite was detected after a time largely more where the equilibrium state was reached. The experiment to be performed when it will be possible, with the ESRF beam line perfectly aligned on the CDW satellite in a sample kept a long enough time at fixed $T^*<T_{CDW1}$, to warm the sample above $T_{CDW1}=336$ K and cool it down again up to $T^*$ and remeasure immediately the satellite profile to see if there are any changes, such a possible time and/or temperature dependence of the rotation of the $Q$ vector, hoping that the temperature excursion will not misalign the beam line.

In conclusion we have shown that, when cooled at the rate of a few degree/min through the Peierls phase transition, the CDW ground state of the rare-earth tritelluride TbTe$_3$ is not at the equilibrium, but time dependent. That is reflected by the increase of the threshold electric field for depinning the CDW and initiation of the sliding with an exponential dependence on time. This phenomenon was not observed for quasi-one-dimensional CDWs and appears to be specific to the two-dimensional nature of the RTe$_3$ compound, specifically linked to Te-Te bonding with different bond distances that form the charge modulation. More experiments to fully describe the phenomenon, namely by quenching the sample through $T_{CDW1}$ or by cooling it through $T_{CDW1}$ down to $T^*$ under an electric field larger the threshold field at $T^*$ are underway.

\section{Acknowledgments}

We are thankful to S.A.~Brazovskii, P. Qu\'{e}merais and O. Cepas for useful discussions. The work has been supported by Russian State Fund for the Basic Research (No. 18-02-00295-a). A.V.F. and A.P.O. thank State assignment IRE RAS.

\newpage

\end{document}